# THAT MESSAGE WENT VIRAL?!

# EXPLORATORY ANALYTICS AND SENTIMENT ANALYSIS INTO THE PROPAGATION OF TWEETS.

## Abstract:

Information exchange and message diffusion have moved from traditional media to social media platforms. Messages on platforms such as Twitter have become the default mode of company communications replacing lengthier public announcements and updates. Businesses and organizations have increased their use of Twitter to connect with stakeholders. As a result, it is important to understand the key drivers of successful information exchange and message diffusion via Twitter. We conducted an exploratory analysis on a dataset of over a million Tweets, comprising of over 40,000 lead Tweets, further filtered to over 18,000 Tweets. We identified the most popular messages, and analyzed the tweets on multiple endogenous dimensions including content, sentiment, motive and richness, and exogenous dimensions such as fundamental events, social learning, and activism. We found some interesting patterns and uncovered new insights to help researchers and practitioners better understand the behavior of popular "viral" tweets. We also performed sentiment analysis and present an early stage model to explain tweet performance.

## Introduction:

Much has already been written about the popularity and ubiquitous nature of Twitter. Twitter has a global presence and influences a variety of important issues and events including election results, consumer confidence, brand development, crisis communication, day to day social interactions and much more. From a behavioral analysis standpoint, the efficacy of information influence has been mapped to the nature of the information artifact (Samuel, 2017) and this has meaningful implications for how information is presented, interpreted and used. Past research has studied Tweet phenomena from a variety of perspectives including the analyses of specific users, topics, and markets. Studies have also focused on case scenarios (Pressgrove, et. al.., 2018; Kalsnes, B., & Larsson, A. O., 2017), while examining the type of news propagated (Vosoughi, S., et. al., 2018; Wahba, G., 2017), and the extent of information richness (Samuel &

Pelaez, 2017). However, past studies have lacked integrative frameworks and necessary theoretical abstractions. Hence, despite their intrinsic value, they highlight an urgent need to study viral information through the lens of appropriate theoretical frameworks (Alter, S., 2017).

The present research uses data from Twitter in the form of tweets and their associated attributes. Twitter has around 330 million users, generating more than 500 million tweets a day. 80% of the tweets originate from mobile devices and about 30% of users are active on an average day. Tweets can be viewed as a massive message propagation network phenomenon, which is being leveraged expansively by individuals and organizations (Fuchs 2017). Twitter's reach is global and there is a strong interest in using tweets from thought leaders and influencers, including government leaders (Weller, et. al., 2014; Dubois, et. al., 2014;). Research has also shown that stock returns and volumes are correlated with related stock messages before, during, and after events (Sprenger, et. al., 2014). Twitter is the world's leading microblogging platform, and typifies information flows on social media networks. It is a rich source of high quality data characterized by low cost of acquisition. However, for larger quantities of Twitter data, the expenses pertaining to the process and computational resources necessary to store and process this information can be challenging. One of the ways in which Twitter data is analyzed includes estimating the 'sentiment' associated with a message – past research has used sentiment analysis to observe patterns in stock price movements and observed that the sentiment of tweets was related to the stock market dynamics(Kretinin et al., 2018). We expect our analysis to help us identify valuable insights from exploratory analytics, sentiment analysis and textual analytics. The present research stream intends to explore the many tweet associated variables that have the potential to make tweets go viral, and to identify factors that evoke emotion and interest when new messages or information in the form of tweets become available.

**Literature review**

Multiple studies of sentiment analysis using Tweet data have found an association between emotionally charged messaging and message popularity. Analysis of political messages on Twitter has shown that emotional messages become popular by being retweeted or marked as favorites (Stieglitz & Dang-Xuan, 2013). Viral propagation of a message can also be driven by powerful response sentiment embedded in simple, yet emphatic hashtags such as "Je suis Charlie" (Payne, R., 2018), which quickly spread virally across Twitter and other social media platforms. While the ‘Je suis Charlie’ phenomenon can be viewed as emotionally reactive activism, additional insights can be gained from the ‘ALS Ice Bucket Challenge’, which also became a popular tweet theme. ALS tweets were often accompanied by, or consisted of, videos of people pouring ice water on their heads, or dipping their heads into ice water buckets. Although this had strong visual and emotional impact, it exemplified structured activism through social media, with elements of social learning. Extant research that has analyzed tweets related to technical information have shown that actual information content matters, and new and actionable information, such as tweets related to “alerts, patch, advisory, exploit, and root-cause” (Syed et al., 2018), tends to become more popular. In addition, the popularity of tweets can be driven by commonly known fundamentals, such as stock related news or announcements. An analysis of UBI (Universal Basic Income) tweets has shown that tweets containing unique information tends to become most popular, while tweets with a high emotional value tends to attract new audiences (Hemsley et al., 2018).

Tweets can also become popular based on the “information richness” they display – this includes the use of special characters, pictures, links and visual or semantic enrichment of the core message. Information richness has been shown to be an important cognitive influence (Samuel & Pelaez, 2017) with potential to sway how users respond to tweets. Furthermore, tweets can also be viewed as categories or information modules with specific properties. For instance, stock market tweets, sports tweets, funny tweets, political tweets, official tweets, trivia and other such categories are bounded by the objective of categorization. Information categories, even with the same net information content, can have varying influence on users (Samuel, Holowczak & Pelaez, 2017). This implies that tweet creators may need to pay attention to the actual composition of the message, the category in which it is classified, and the richness of message communicated. Clinical approaches have been used to study tweets by classifying tweets analysis along structural, content, and sentiment dimensions to explain tweet popularity in the context of the number of followers, length of the tweet, and the use of features such as hashtags, live video streaming, and so forth (Jenders et al., 2013). Our initial analysis of tweets shows that Twitter data is rich and contains hitherto unexplored variables that form the basis of our approach towards data collection and analysis.

*Research Objectives*

- Our research objective with the present study is to provide early stage insights into the characteristics of viral tweets.
- We also aim to identify key variables that can be associated with viral tweets.

Despite these recent advancements into the explanation of information exchange and diffusion within social networks, there still exist many gaps within the extant literature. There is a fair degree of ambiguity surrounding the definitions used to frame "viralness", and numerous tweet associated variables available within Twitter datasets are yet to be rigorously analyzed. Hence, the goal of this research is to explore the potential factors that may drive "viral" tweets, as well as render additional research questions for future contribution. We conduct an exploratory analysis that identifies the various observables that are reflective of a "viral" tweet, filter the variables that are associated with the observables which could be argued to be reflective of "viralness", and how such variables influence said observables. We conclude with an early stage model, and present unique findings that provide additional understanding of viral tweets phenomena.

**Methodology**

We used two twitter accounts to download Twitter Status IDs, Twitter User IDs, and number of recounts, among a number of variables. At our starting point of our analysis we collected over a million tweets with over eighty variables. We used MySQL to store the raw Twitter data and 2 Different R scripts were written, which leveraged the `rtweet` package and the `RMySQL` packages.

*The Twitter Application Programming Interface (API)*

The goal of this research is to conduct exploratory analysis, and hence, for understanding the dimensions of "viralness", and the various aspects that may affect this, we explore the various components and attributes of a Tweet. The Twitter API is a collection of web-service functions that can be accessed via HTTP Requests. These types of calls and APIs have recently become popularized by many social media platforms as a way to provide programmatic access to organizations and users alike into the Twitter databases. Typically, these APIs take in well-structured HTTP Requests, and returns either an XML or a JSON formatted response. Across multiple programming languages, various packages and libraries exist to handle much of the boilerplate procedures to properly structure the HTTP Requests and return the information that Twitter has provided in JSON/XML format. In this study, we leveraged R and the rtweet package. This package allows for the ease of use of interaction between the user/programmer and the Twitter API. Functions within the package handle the structuring of the HTTP Requests. Furthermore, they are able to structure the responses given within the JSON format into a simple data frame, which is one of the fundamental forms of data structures that R provides. This allows for the analysis of Twitter data to be more streamlined, and frees the user from more traditional data processing and data preparatory tasks.

*Tweet Characteristics*

When a request is made to the Twitter API, by default, Twitter will provide over 80 different characteristics regarding the tweets requested. We analyze these 83 variables to gain insights into the various information and social dynamics between and about users on Twitter.

| Tweet Variable Name | Description | Type |
|---|---|---|
| user_id | A unique identification number of the twitter user. | User |
| status_is | A unique identification number for the specific tweet, retweet, or quote. | Tweet |
| created_at | The timestamp of creation for the tweet. | Tweet |
| text | The actual text of the tweet. This will contain text, hashtags, and urls. | Tweet |
| source | The type of device from which the tweet was generated. Many possibilities can be assigned here. For example, "Twitter for iPhone" is one example. | Tweet |
| display_text_width | The number of characters of the tweet that can be seen by the user. | Tweet |
| favorite_count | The number of "likes" for this specific tweet. | Tweet |
| retweet_count | The number of times this tweet was retweeted by other users. | Tweet |
| hashtags | A list of hashtags that were used within the tweet. | Tweet |
| media_type | An indicator to determine if the tweet had a photo embedded within it. | Tweet |
| lang | The language of the tweet. | Tweet |
| location | The location of the user account. | User |
| follwers_count | The number of followers that this user currently has. | User |
| friends_count | The number of users that this user follows. | User |
| listed_count | The number of public lists for which this user is a member. | User |
| statuses_count | The total number of tweets,retweets, or quotes that this user has posted. | User |
| favourites_count | The total number of tweets that this user has "liked". | User |
| account_created_at | The timestamp of the creation of the user's account. | User |
| verified | An indicator variable to determine if Twitter has verified this user's account information. | User |
| account_land | The language of the user. | User |

Table 1: Selected variables within a processed response from the Twitter API.

These tweet characteristics are categorized into two types of meta-information: tweet-specific information and user-specific information. Table 1 offers a summary of the most relevant tweet and user characteristics that we found important in relation to our research questions and objectives. Given that our research was primarily exploratory, we first defined our intended population of interest, designed a data-gathering process, designed code to clean the data once it was fully gathered, and then ran an exploratory data analysis. The following subsections profile these steps in detail. We were primarily interested in gathering data consisting of tweets of which themselves were retweeted. In order to ensure that we had proper levels of heterogeneity within our data, and were able to identify tweets that were more "viral" and those that were not, we gathered tweet information that had a wide variance in retweet and favorite count. However, to remove some potential data analysis problems (mainly the high levels of zero-inflation in our data set), we made the decision to ignore tweets that had no retweets and favorites.

*Data Collection and Cleaning*

The data gathering process was implemented within R, which leveraged the `rtweet` and the `RMySQL` packages. The collection effort was split into two different scripts. The first script implemented the data gathering of the retweet keyword search, and the second script implemented the data gathering of the trending topic tweets. Each script sent a query to the Twitter API via `rtweet` with an upper limit of 15,000 tweets for every download cycle. The query only focused on tweets that were generated from within the USA, and only the top 25% of tweets were gathered to help filter near-zero interest tweets. Once the tweets were downloaded, their user id and status id were stored within a `MySQL` Database. All other information were not extracted for the sake of saving physical memory and cost. These two scripts were saved and both ran using two different Twitter accounts. The scripts were run daily, every 15 minutes as a CRON job on an Amazon Amazon EC2 t2.large Ubuntu Server instance, which had 2 virtual processors and 8.0Gib (≈ 8.58GB) of RAM. The MySQL Database was running on a server-less instance leveraging Amazon RDS. This allowed for the data to be placed in a centralized area, separate from the machine, so that in the event of a machine error or accidental shutdown, the data would still be intact. The scripts were ran throughout the entire month of November, 2018. Subsequently an Amazon EC2 m5.24x large instance was started. This machine had 96 processors with 384 Gib of RAM (≈412 GB), and was used to gather the remaining raw twitter data. The final data was processed and stored as a .rds, which is a compressed file format for storing R objects. The raw data file was approximately 13.5GB, but the rds compression reduced this down to ≈2GB.

Tweets were filtered from the larger dataset based on the tweets that were indicated to be English and the account language to be English. The top 75% of tweets based on retweets were then filtered, and this was done to ensure that the data we were analyzing had retweets. The source variable was then cleaned, and the values were combined to be either “Mobile”, “Desktop” or “Other” based on keywords related to these categories. The following scheme was used: Tweets were classified to be posted from a “Mobile” device if the source contained the following keywords: "iPhone","iOS", "BlackBerry", "Tablets", "Android", "Phone","iPad", "Mobile". Tweets were classified to be posted from a “Desktop” device if the source contained the following keywords: "Windows", "Mac", "Web Client". All other tweets were simply classified as “Other”. Tweets that contained text (other than a hyperlink) were filtered from tweets that contains no text, since one of the primary objectives of this research was to analyze tweets based on the content of the text.

*Data Preparation and Variable Refinement*

Part of our exploratory analysis involved the use of lexicons to compute overall sentiment of a tweet. A sentiment is a score that represents the overall emotion of the tweet. In our research, sentiments are the same as total polarity. Generally speaking, in order to compute the polarity of text information, each word is assigned a number according to a given lexicon. The lexicon is then used to compute the overall polarity of the statement by simply adding the polarity values in the lexicon together for the words that appear in the text. Text that is not in the lexicon is assumed to have a polarity of 0, hence not affecting the overall score. Many different lexicons are available for this type of analysis. In R, nine lexicons are available and we used all nine to compute the sentiment of every tweet. The text of the tweet was used to find the polarity, and polarities were computed for all the tweets using each of the lexicons. A total of nine sentiment variables were computed and appended to the overall dataset. The following lexicons were used to compute the 9 different sentiment ratings in R:

```
hash_sentiment_huliu, hash_sentiment_jockers,
hash_sentiment_jockers_rinker, hash_sentiment_loughran_mcdonald,
hash_sentiment_nrc, hash_sentiment_senticnet,
hash_sentiment_sentiword, hash_sentiment_slangsd and
hash_sentiment_socal_google
```

In addition, the created date was split using R's base functionality into the day of the week, the day of the month, the month, the year, and the hour that the tweet was created.

Furthermore, an account age was computed by taking the difference in date between the time stamp the tweet was created and the timestamp the account was created. This variable is in terms of the number of days between the day the account was created and the day the tweet was created.

## Exploratory Analytics

The first part of our analysis involved an exploratory factor analysis (EFA) on all of the numerical variables in the data set. Ideally, we would like to identify distinct constructs within the data. The numerical variables were extracted from the data set and moved into it's own data frame. We leveraged R's psych package in order to conduct this type of analysis. The EFA was conducted by using a minimum residual method with a varimax rotation. The results of the EFA indicated an overall fit with the RMSEA <0.05 and the Tucker Lewis Index >0.95, which indicates an overall good fit of the analysis.

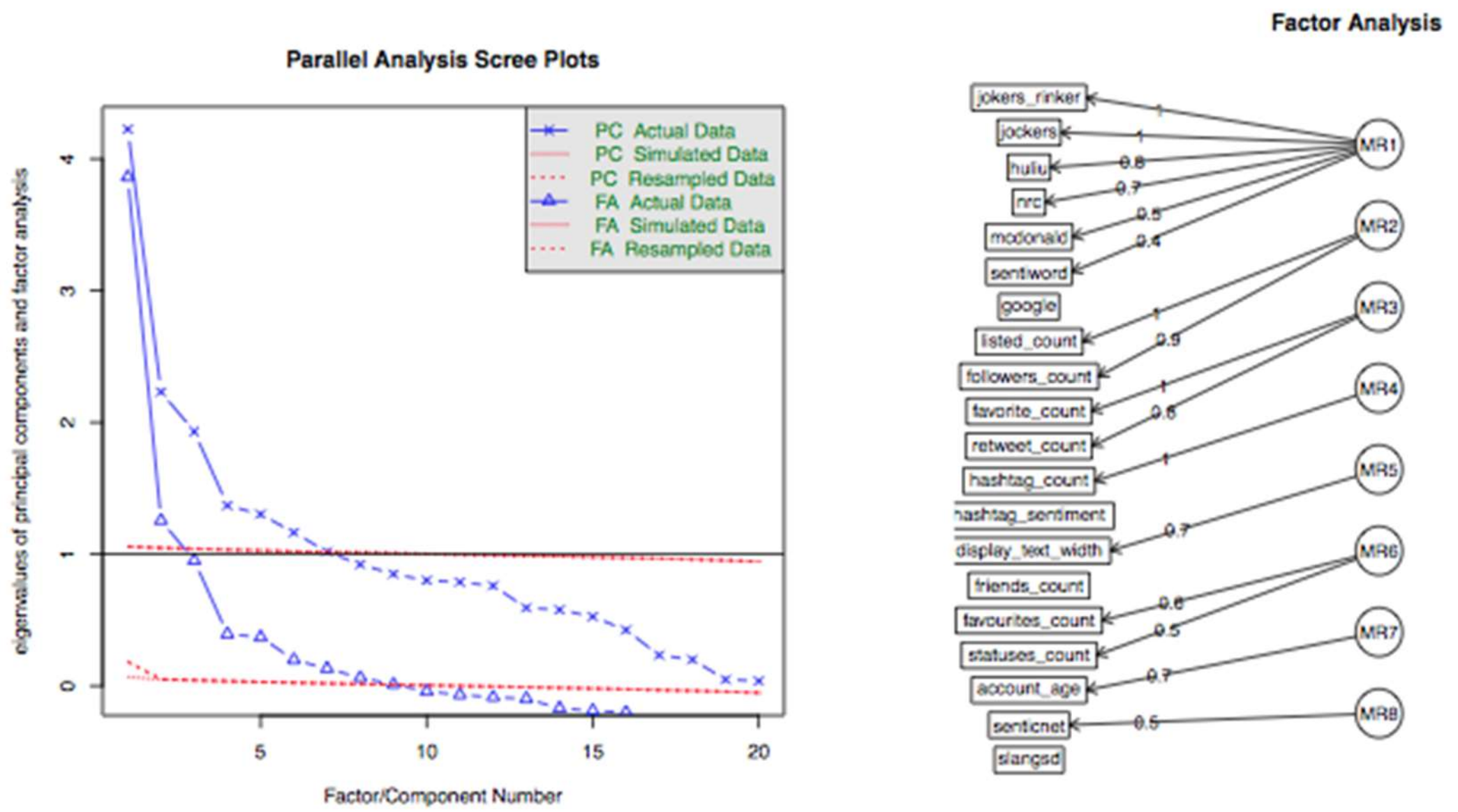

Scree Plot of Numerical Variables

Factor Analysis of Numerical Variables

Figure 1: Exploratory Factor Analysis

We can see from Figure that there were a total of 8 factors. The raw sentiment variables had loaded on 2 of the 8 factors. Listed count and followers count loaded on the same factor. Favorite count (which indicates the total number of "likes" of the specific tweet) and retweet count loaded on the same factor. Hashtag count loaded on it's own factor, as did the display text width and the account age, respectively. Last, the user's favourites count (which indicates the total number of tweets the user "liked") loaded on the same factor as the number of statuses.

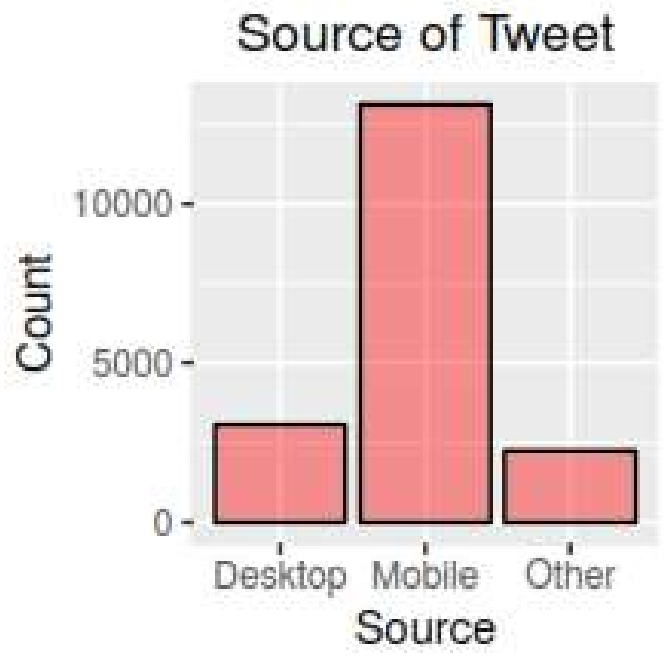

Figure 2A.

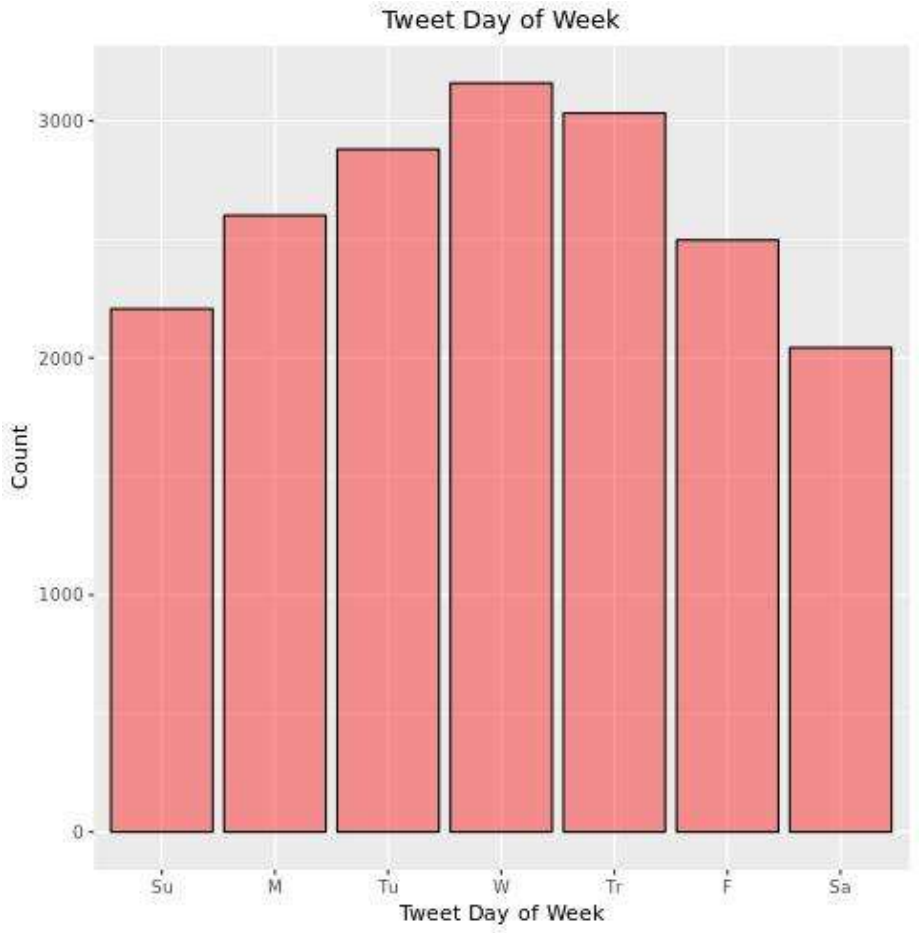

Figure 2B.

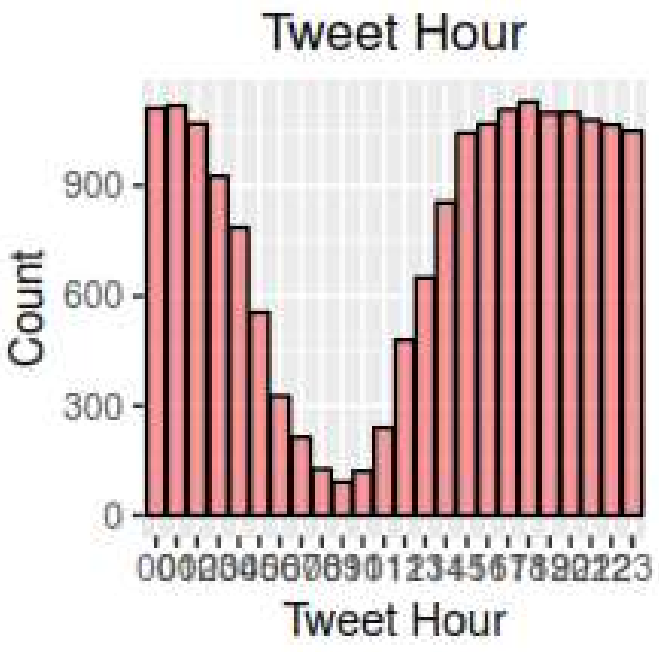

Figure 2C.

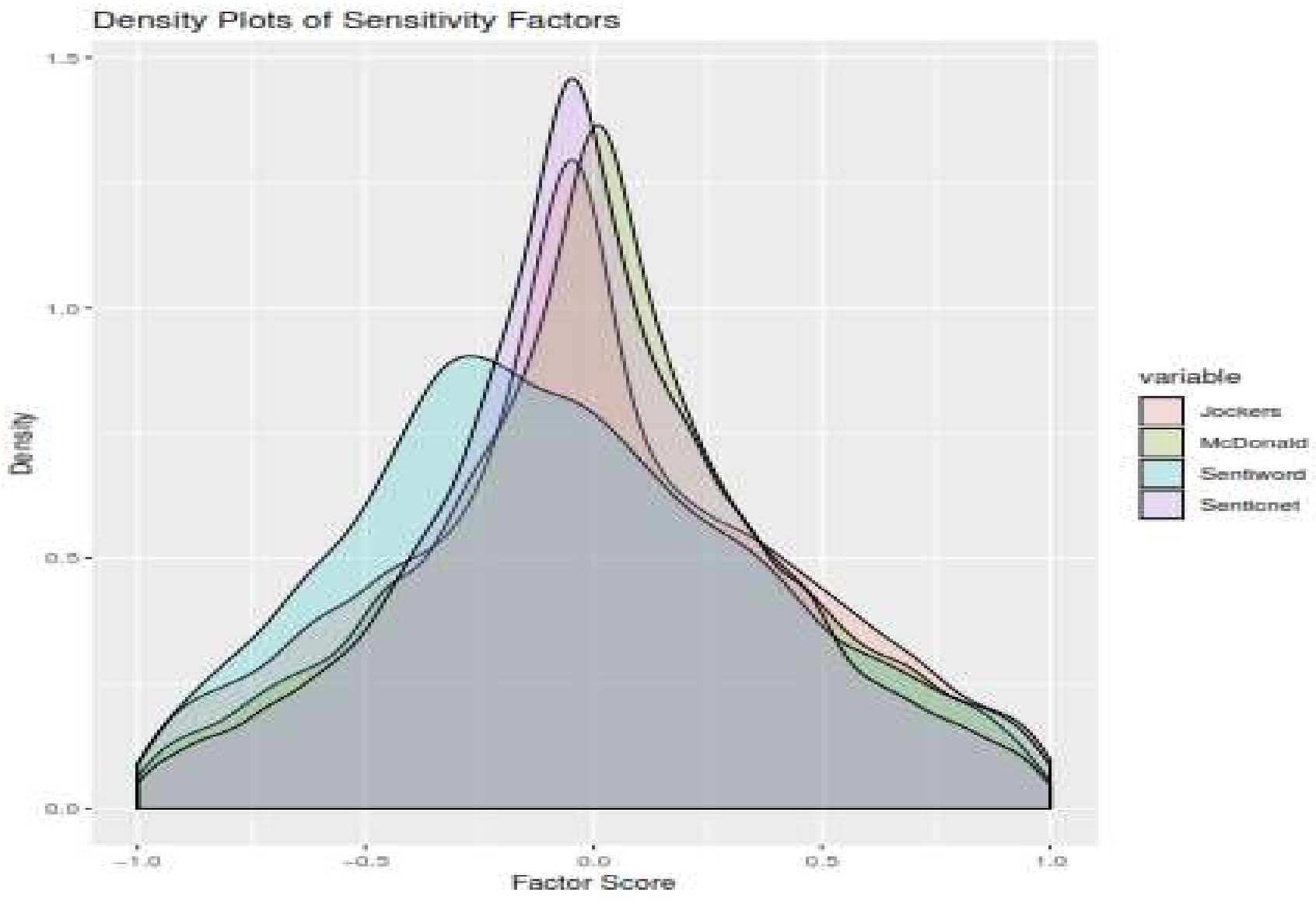

Figure 2D.

In order to reduce the number of variables, and given some initial evidence from the EFA, we decided to conduct a separate EFA on only the nine sentiment variables so that we would be able to reduce the number of variables we will use in the remainder of our analysis. Results of this EFA indicate an overall fit with the RMSEA=0.06 with the Tucker Lewis Index >0.95, which indicates a good fit of the model.

Table 2: Summary Statistics for the Tweet Variables

| Statistic | N | Mean | St. Dev. | Min | Pctl(25) | Pctl(75) | Max |
|---|---|---|---|---|---|---|---|
| Jockers | 18,420 | 0.000 | 0.884 | −15.076 | −0.377 | 0.385 | 14.047 |
| McDonald | 18,420 | −0.000 | 0.673 | −7.895 | −0.281 | 0.298 | 8.408 |
| Sentiword | 18,420 | −0.000 | 0.746 | −5.093 | −0.419 | 0.339 | 13.614 |
| Senticnet | 18,420 | 0.000 | 0.559 | −7.092 | −0.246 | 0.250 | 14.091 |
| display_text_width | 18,420 | 131.331 | 87.428 | 4 | 56 | 208 | 468 |
| favorite_count | 18,420 | 53,530.650 | 90,771.310 | 422 | 13,589 | 55,867 | 4,518,186 |
| retweet_count | 18,420 | 18,596.850 | 40,615.700 | 3,683 | 5,415.8 | 18,204.2 | 3,317,687 |
| followers_count | 18,420 | 2,867,175.000 | 10,983,453.000 | 15 | 9,856.8 | 756,371 | 103,532,188 |
| friends_count | 18,420 | 21,427.190 | 95,160.310 | 0 | 387 | 3,608.2 | 1,649,521 |
| listed_count | 18,420 | 7,008.633 | 21,543.580 | 0 | 91 | 3,743.2 | 229,879 |
| statuses_count | 18,420 | 36,850.020 | 65,947.170 | 0 | 4,443 | 42,049 | 1,871,122 |
| favourites_count | 18,420 | 25,892.170 | 57,546.450 | 0 | 1,432 | 25,854 | 1,209,032 |
| account_age | 18,420 | 2,235.632 | 1,172.548 | 0 | 1,196 | 3,387.2 | 4,477 |
| hashtag_count | 18,420 | 0.179 | 0.665 | 0 | 0 | 0 | 15 |
| hashtag_sentiment | 18,420 | 0.031 | 0.513 | −9 | 0 | 0 | 9 |

Figure 3 indicates the sentiment variables loaded onto four different factors. We identified four distinct sentiment systems, which we identify as Jockers, McDonald, Sentiword, and Senticnet. The corresponding factor scores for each tweet for each of these sentiment factors were extracted and appended to the primary data frame which was used for the remainder of the data analysis.

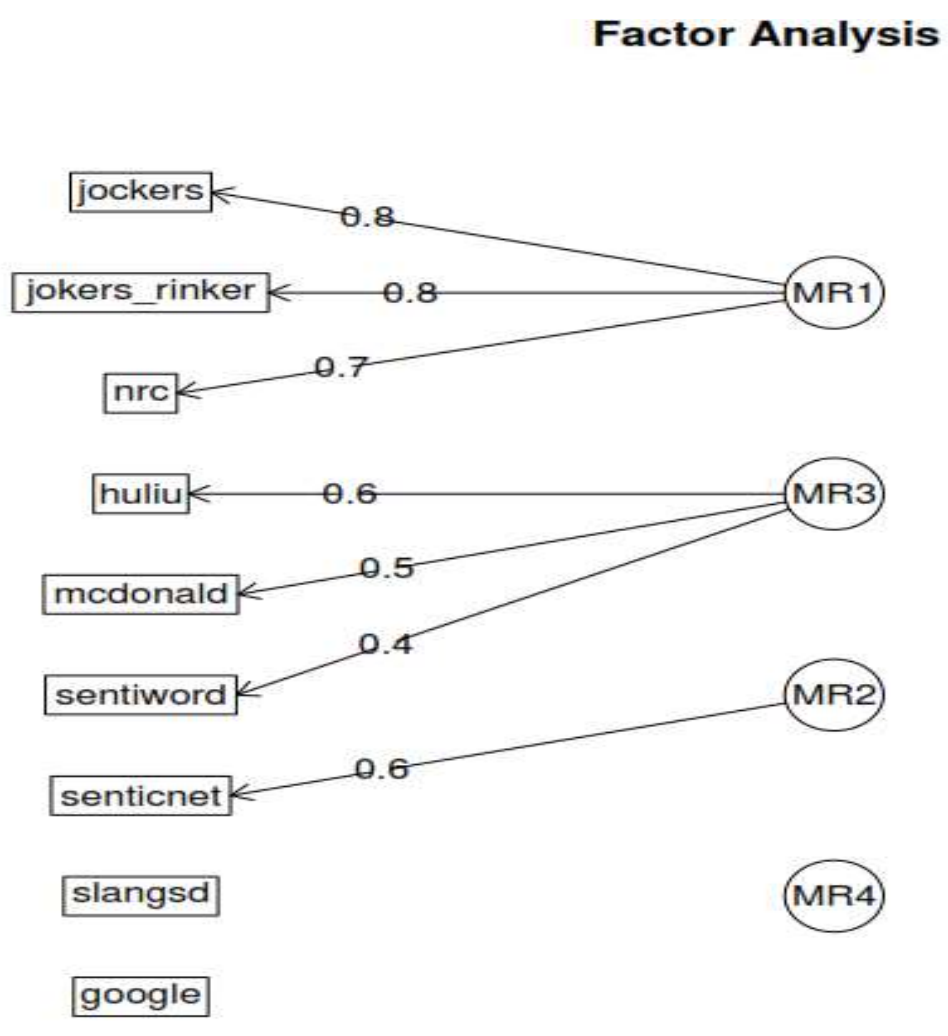

Figure 3.

A descriptive analysis was conducted, which reported the measures of central tendency, dispersion, and other moment information. The results for the numerical variables are shown in Table 2. Our final data set for analysis comprised of 18420 unique and distinct most popular tweets. Almost all of the tweets in the data set were created within November, during the time we were collecting information. Most tweets tend to occur at the beginning to the middle of the month. In addition, there are significantly less number of tweets during the middle of the day. Our data has fewer verified accounts than non-verified, and we have more tweets without a photo. We also notice that most of the tweets originated from a mobile device rather than a more traditional desktop. Last, we notice that the highest tweet activity seems to occur from Tuesday to Thursday, during the heart of the work week. This would seem to indicate that most Tweets are posted and retweeted during the middle of the work week.

Some of the plots for these categorical variables are show in Figure 2. Table 3 reports the correlation table for the numerical variables. We notice that the retweet count, as well as the favorite count, are significantly correlated with most of the other variables. We also notice a potential for multicollinearity, since the sentiment Jockers is correlated with all other variables other than retweet count and friends count. The hashtag count is also correlated with all but 5 of the variables. Observing the plots of the retweet count against the other variables reveals some potentially interesting associations. In addition, we can see from Figure that the McDonald and Sentinet sentiment scores are very similar to Jockers and Sentiword, respectively. Hence, to further simplify our model, we will only consider the two most different sentiment scores: Jockers and Sentiword.

| | 1 | 2 | 3 | 4 | 5 | 6 | 7 | 8 | 9 | 10 | 11 | 12 | 13 | 14 |
|---|---|---|---|---|---|---|---|---|---|---|---|---|---|---|
| 1.Jockers | | | | | | | | | | | | | | |
| 2.McDonald | 0.32* | | | | | | | | | | | | | |
| 3.Sentiword | 0.01* | 0.12* | | | | | | | | | | | | |
| 4.Senticnet | 0.46* | 0.41* | -0.03* | | | | | | | | | | | |
| 5.display_text_width | -0.02* | -0.11* | 0.31* | 0.02* | | | | | | | | | | |
| 6.favorite_count | 0.01* | 0.02* | -0.03* | -0.01 | -0.12* | | | | | | | | | |
| 7.retweet_count | 0.01 | 0.02* | -0.04* | -0.01 | -0.11* | 0.80* | | | | | | | | |
| 8.followers_count | 0.06* | 0.06* | 0.07* | 0.10* | 0.09* | 0.17* | 0.13* | | | | | | | |
| 9.friends_count | 0.00 | -0.01 | 0.04* | 0.01 | 0.14* | 0.01 | 0.01 | 0.07* | | | | | | |
| 10.listed_count | 0.05* | 0.05* | 0.08* | 0.08* | 0.11* | 0.16* | 0.12* | 0.95* | 0.10* | | | | | |
| 11.statuses_count | -0.02* | 0.00 | -0.01 | -0.01 | 0.00 | -0.07* | -0.04* | 0.01 | 0.09* | 0.05* | | | | |
| 12.favourites_count | -0.03* | -0.02* | -0.02* | -0.03* | 0.03* | -0.01 | 0.00 | -0.10* | 0.10* | -0.11* | 0.30* | | | |
| 13.account_age | 0.02* | -0.01 | 0.09* | 0.05* | 0.21* | -0.06* | -0.07* | 0.23* | 0.08* | 0.28* | 0.24* | -0.03* | | |
| 14.hashtag_count | 0.05* | 0.03* | 0.04* | 0.08* | 0.14* | -0.07* | -0.04* | 0.00 | -0.01 | 0.00 | 0.00 | 0.00 | 0.06* | |
| 15.hashtag_sentiment | 0.08* | 0.05* | 0.03* | 0.05* | 0.02* | -0.01 | 0.00 | 0.01 | 0.00 | 0.01 | 0.01 | 0.00 | 0.02* | 0.20* |

Table 3: Correlation Table for Numerical Variables (* $p \leq 0.05$)

We observed from plots not shown in this paper that the sentiment variables and the retweet count appear to be within a curvilinear association. The remaining variables appeared linear, as also suggested by the correlation table. Hence, for the sentiment variables, we will square each one and include the squared version in our overall model. The final variables that we will study in our regression analysis will be the Jockers and Sentiword sentiment factors with their squared terms, the text display width, the favorite count, the followers count, the friends count, the listed count, the statuses count, the favourites count, the account age, the hashtag count, the hashtag sentiment, the square of the hashtag sentiment, the month, the day of the week, the hour, the verified indicator, the source, and the inclusion of a picture.

**Regression Analysis**

We found from our EFA that the retweet count and the favorite count loaded on the same factor. It could be argued that this factor could represent a "viralness" or "popularity" construct. For this study, we will focus on the retweet count and use the favorite count as a control variable. In our regression analysis, we seek a model that can explain the retweet count for a given set of independent variables. The dependent variable in this model will be the retweet count. However, since this variable represents a count, the linear regression model with the ordinary least squares estimate will most likely lead to many problems, including (1) misspecification, (2) inconsistent or biased estimates of the parameters and (3) inefficient estimates of the parameters. Hence, we must take an alternative approach. When the dependent variable is a count variable, (Cameron and Triveddi, 2013) recommend using a Poisson Regression. Poisson regression with independent variables xi seeks to find the probability distribution for the dependent count variable by estimating the average count (i.e. the incidence rate). Recall that the only parameter for the Poisson distribution is the average number of arrivals λ. In our model, we seek to estimate the model λ = E[yi|xi] = eβ`xi for a given individual i, which would enable us to find:

$$\begin{aligned} P(y_i = k|\boldsymbol{x}_i) &= \frac{\lambda^k}{k!}e^{-\lambda} \\ &= \frac{(e^{\boldsymbol{\beta}^T\boldsymbol{x}_i})^k}{k!}e^{-e^{\boldsymbol{\beta}^T\boldsymbol{x}_i}} \end{aligned} \tag{1}$$

where β`xi = β0 + β1 xi,1 + ... + βn xi,n. However, one major drawback regarding the Poisson Distribution is that it must enforce the moment constraints of E[yi|xi] = Var[yi|xi], by definition of Poisson. This design can be quite troublesome, since count data tends to already be heteroscedastic, and hence the estimate for the Poisson model would most likely be inefficient (Cameron and Treveddi, 2013; Greene, 2003). In order to correct for this, we will employ a negative binomial model. The negative binomial model is a generalization of the Poisson regression, and has been used in the analysis of Twitter data in the extant literature (Syed et al. 2018; Stieglitz & Dang-Xuan, 2003). The difference between these models is that the variance and the mean are not restricted to equality. Rather, the data is modeled by allowing the variance to differ non-linearly using a dispersion parameter. Furthermore, the use of a negative binomial model would help remove much of the heteroscedasticity by design, since much of this variation is captured within the negative binomial distribution itself, as well as how the error terms are modeled. Since the Poisson and Negative binomial do not have closed-form first order conditions for their respective log-likelihood functions, many of these estimators can only be solved through iterative optimization algorithms (such as Newtons Method leveraging maximum likelihood) (Cameron and Treveddi, 2013). An alternative approach would be to estimate the model leveraging generalized method of moments (GMM), however, these become more difficult to accomplish when endogeneity enters the equation (Wooldridge, 2010). In the case of our study design, we should expect the presence of endogeneity.

We have chosen to conduct a cross-sectional study rather than a panel-data study to ascertain initial information on the potential predictors of retweet count within the static sense. Our time frame for the data collection was also fairly limited, and hence we would not have a large enough time-span to study the time-dependent effects on our variables. The primary limitation of our design, however, is the introduction of potentially endogenous relationships. For example, an argument can be made that the number of followers of a user will lead to a higher retweet count. Conversely, the number of retweets could lead the user to obtain a larger following, since their own user profile becomes more exposed as a result of the higher retweets. It would be ideal to have control information on the number of followers (among others) the user had before posting the tweet (which our data do not provide), then the number of retweets has a probability of being in an endogenous relationship with the number of followers. However, based on heuristic observation, it is fairly evident that most Twitter users would be more likely to follow a Twitter user or account and then retweet messages posted to the account rather than to follow based on retweeting. (Wooldridge, 2010) recommends applying a modified version of the 2 stage least squares (2SLS) approach to estimate our model. First, the potentially endogenous variables are identified. The first stage of the regression is to regress the exogenous variables over the endogenous variables. The residuals from this regression are then used within the full model specification with the speculated endogenous variables. A Wald-Test can then be conducted to determine if the coefficient on the residual is 0. If it is, then there was little evidence of endogeneity in the first place, and this approach would not offer any advancement in consistency and unbiasedness. In order to carry out this procedure in a more simplistic fashion, we need to identify an instrument within our study. We notice that the follower count and the listed count are highly correlated with each other (ρ = 0.95).

Furthermore, it would be difficult to find an argument to suggest that the listed count would be affected by the number of retweets. If a user is a member of a list, then it is possible that members of list will notice the user's tweet, and hence retweet it. However, the list is it's own entity within Twitter, and it would be a much longer line of reasoning to suggest that if a user's tweet were highly retweeted, and the users who retweeted are members of different groups, then the user that posted the tweet would be motivated to now follow the list. That is, the direction of causality is most likely that the list count will aid in the retweet count, but the number of retweets will not necessarily lead to a change in the list count. Therefore, it would appear to be unlikely that the listed count variable would be correlated with the error term, and hence, by definition, would be an instrument for the followers count. Therefore, our regression analysis will focus on the following empirical model for the first stage:

$$\begin{aligned}\lambda_{retweet_count,stage_1} &= E[retweet_count_i|\boldsymbol{x_i}] \\ &= \exp(\beta_0 + \beta_1 Jockers_i + \beta_2 Jockers_i^2 + \beta_3 Sentiword_i + \beta_4 Sentiword_i^2 \\ &+ \beta_5 account_age_i + \beta_6 display_text_width_i + \beta_7 verified_i \\ &+ \beta_8 source_i + \beta_9 media_type_i + \beta_{10} favorite_count_i + \beta_{11} created_month_i \\ &+ \beta_{12} created_day_i + \beta_{13} created_hour_i + \beta_{14} friends_count_i \\ &+ \beta_{15} favourites_count_i + \beta_{16} statuses_count + \beta_{17} hashtag_count \\ &+ \beta_{18} hashtag_sentiment + \beta_{19} listed_count + \epsilon_i)\end{aligned} \tag{2}$$

The second step in the regression is to find the residuals from the first stage, $\hat{\epsilon_i}$, and include these in the model as a predictor for our final empirical model with the endogenous relationship:

$$
\begin{aligned}
\lambda_{retweet_count} &= E[retweet_count_i|\boldsymbol{x_i}] \\
&= \exp(\beta_0 + \beta_1 Jockers_i + \beta_2 Jockers_i^2 + \beta_3 Sentiword_i + \beta_4 Sentiword_i^2 \\
&+ \beta_5 account_age_i + \beta_6 display_text_width_i + \beta_7 verified_i \\
&+ \beta_8 source_i + \beta_9 media_type_i + \beta_{10} favorite_count_i + \beta_{11} created_month_i \quad (3) \\
&+ \beta_{12} created_day_i + \beta_{13} created_hour_i + \beta_{14} friends_count_i \\
&+ \beta_{15} favourites_count_i + \beta_{16} statuses_count + \beta_{17} hashtag_count \\
&+ \beta_{18} hashtag_sentiment + \beta_{20} followers_count + \beta_{21}\hat{\epsilon}_i + u_i)
\end{aligned}
$$

To further simplify the model, we took an iterative process to determine if we can remove some of the predictor variables that did not add any meaningful contribution to the empirical model, without sacrificing our exploratory analysis. We did so by estimating nested OLS models and subsequently running a likelihood ratio test for each iteration. This test is appropriate for model comparison given that we have nested models under comparison (Greene,2003). Interestingly enough, the results indicated that that variables of friends, listed, and text width did not add any meaningful contribution to the explanatory power of the model. However, we made the decision to leave these variables within the model for two reasons. First, a theoretical justification, due to some of the extant literature, can be found as to why these factors may influence the retweet count. Second, these variables are of interest to us, and we are curious to determine how they impact the probability of retweets, despite their contribution being minimal.

In addition to these results, the day of the month was removed, despite a likelihood ratio test indicating a statistically significant difference in the model specifications. The reasoning behind our choice to remove it was (1) while the likelihood ratio test indicated a statistically significant difference, the p-value was not as low ($p \approx 0.03$) as other tests, and (2) there were 31 categories in this variable, which would equate to having 30 additional coefficients in the model. Hence, we decided to remove this variable from the model, despite it adding some minor level of explanatory power. Next, the remaining categorical variables were removed one-by-one and compared using the likelihood ratio test. All tests showed that they do indeed add significant levels of explanatory power to the overall model. The final model is hence as follows:

$$
\begin{aligned}
\lambda_{retweet_count,stage_1} &= E[retweet_count_i|\boldsymbol{x_i}] \\
&= \exp(\beta_0 + \beta_1 Jockers_i + \beta_2 Jockers_i^2 + \beta_3 Sentiword_i + \beta_4 Sentiword_i^2 \\
&+ \beta_5 account_age_i + \beta_6 display_text_width_i + \beta_7 verified_i \\
&+ \beta_8 source_i + \beta_9 media_type_i + \beta_{10} favorite_count_i + \beta_{11} created_month_i \\
&+ \beta_{12} created_hour_i + \beta_{13} friends_count_i \\
&+ \beta_{14} favourites_count_i + \beta_{15} statuses_count + \beta_{16} hashtag_count \\
&+ \beta_{17} hashtag_sentiment + \beta_{18} listed_count + \epsilon_i)
\end{aligned}
\tag{4}
$$

$$
\begin{aligned}
\lambda_{retweet_count} &= E[retweet_count_i|\boldsymbol{x_i}] \\
&= \exp(\beta_0 + \beta_1 Jockers_i + \beta_2 Jockers_i^2 + \beta_3 Sentiword_i + \beta_4 Sentiword_i^2 \\
&+ \beta_5 account_age_i + \beta_6 display_text_width_i + \beta_7 verified_i \\
&+ \beta_8 source_i + \beta_9 media_type_i + \beta_{10} favorite_count_i + \beta_{11} created_month_i \\
&+ \beta_{12} created_hour_i + \beta_{13} friends_count_i \\
&+ \beta_{14} favourites_count_i + \beta_{15} statuses_count + \beta_{16} hashtag_count \\
&+ \beta_{17} hashtag_sentiment + \beta_{19} followers_count + \beta_{20} \hat{\epsilon_i} + u_i)
\end{aligned}
\tag{5}
$$

**Results**

In all models (OLS, Poisson, and NB), we notice there the number of followers and the retweet count were indeed endogenous. For all three models, the coefficients on the residual terms, respectively, were highly significant ($p<0.05$). The different sentiment scores were also found to affect the retweet counts differently. The Jockers sentiment was found to be in a positive quadratic relationship, while the Sentiword was found to be in a negative quadratic relationship. This would indicate that some sentiment lexicons would transcend higher negative/positive sentiments to be highly effective in rendering increases in retweet counts, while other systems may have the opposite effect (which is more in line with intuition). Surprisingly, across all models, the account age was found to be in a negative linear association with the retweet count. This would indicate that the older the twitter account, the less likely it would be to have original tweets retweeted. This seems to counter our basic intuition. The display text width, which is a measure of visual presence of the tweet itself, is in a negative linear association with retweet count. This would indicate that longer worded tweets are less likely to be retweeted. While the OLS model indicated that verified accounts tend to be retweeted more, this outcome is most likely due to the heteroscedasticity. In the more appropriate models of Poisson and NB, we find that this relationship is actually negative. That is, verified accounts are less likely to be retweeted.

| | Dependent variable: | | | | | |
|---|---|---|---|---|---|---|
| | retweet_count | | | | | |
| | OLS | | Poisson | | negative binomial | |
| | (1) | (2) | (3) | (4) | (5) | (6) |
| Jockers | −414.1521000000000000** | −413.2026000000000000*** | 0.0071665290000000*** | 0.0002781768000000*** | −0.0034149190000000 | −0.0066935720000000 |
| | (203.5318000000000000) | (0.8078325000000000) | (0.0000671269600000) | (0.0000676948000000) | (0.0115368600000000) | (0.0106918100000000) |
| Jockers2 | −0.6242209000000000 | −0.7047026000000000*** | 0.0014185180000000*** | 0.0006013437000000*** | 0.0008635486000000 | 0.0008218229000000 |
| | (54.3785400000000000) | (0.2157603000000000) | (0.0000170492100000) | (0.0000182470400000) | (0.0030823550000000) | (0.0028555690000000) |
| Sentiword | −278.7907000000000000 | −280.6876000000000000*** | −0.0161845500000000*** | −0.0121567800000000*** | −0.0089135430000000 | −0.0061359830000000 |
| | (271.3545000000000000) | (1.0765820000000000) | (0.0000866829900000) | (0.0000869886800000) | (0.0153812700000000) | (0.0142484800000000) |
| Sentiword2 | −16.7738800000000000 | −16.3989300000000000*** | 0.0020006930000000*** | 0.0020119610000000*** | −0.0002524719000000 | −0.0006244841000000 |
| | (80.9941800000000000) | (0.3213562000000000) | (0.0000254136000000) | (0.0000251744700000) | (0.0045910360000000) | (0.0042531410000000) |
| account_age | −1.2404830000000000*** | −1.2496460000000000*** | −0.0000311805200000*** | −0.0000279546600000*** | −0.0000377222500000*** | −0.0000377684000000*** |
| | (0.1884565000000000) | (0.0007455984000000) | (0.0000000577081600) | (0.0000000576363500) | (0.0000106823300000) | (0.0000098679550000) |
| display_text_width | 0.3152791000000000 | 0.2999773000000000*** | −0.0011065710000000*** | −0.0010170420000000*** | −0.0005192909000000*** | −0.0004953214000000*** |
| | (2.4032700000000000) | (0.0095349160000000) | (0.0000007930209000) | (0.0000007943853000) | (0.0001362253000000) | (0.0001261947000000) |
| verified | 1,830.2440000000000000*** | 1,801.1830000000000000*** | −0.3173326000000000*** | −0.3347928000000000*** | −0.1062576000000000*** | −0.0929413200000000*** |
| | (459.2268000000000000) | (1.8109750000000000) | (0.0001500729000000) | (0.0001496343000000) | (0.0260304800000000) | (0.0239685500000000) |
| sourceMobile | −1,661.8910000000000000*** | −1,655.3360000000000000*** | 0.1594596000000000*** | 0.1591961000000000*** | 0.0466449800000000 | 0.0441862600000000* |
| | (510.8531000000000000) | (2.0291870000000000) | (0.0001740840000000) | (0.0001755263000000) | (0.0289569100000000) | (0.0268563900000000) |
| sourceOther | 272.8282000000000000 | 250.0624000000000000*** | 0.0909101200000000*** | 0.1085698000000000*** | −0.0129141900000000 | −0.0037291940000000 |
| | (687.7941000000000000) | (2.7285890000000000) | (0.0002472281000000) | (0.0002480653000000) | (0.0389865800000000) | (0.0361129300000000) |
| media_typephoto | 3,054.1640000000000000*** | 3,059.9960000000000000*** | 0.0062699530000000*** | 0.0074070130000000*** | 0.0558180900000000** | 0.0451509400000000** |
| | (399.3365000000000000) | (1.5845250000000000) | (0.0001208696000000) | (0.0001210160000000) | (0.0226356900000000) | (0.0209733400000000) |
| favorite_count | 0.3558628000000000*** | 0.3558048000000000*** | 0.0000019235670000*** | 0.0000021392610000*** | 0.0000090723190000*** | 0.0000091050450000*** |
| | (0.0020727250000000) | (0.0000082137790000) | (0.0000000002167582) | (0.0000000002303112) | (0.0000001174866000) | (0.0000001087080000) |
| friends_count | 0.0013372090000000 | 0.0012582890000000*** | −0.0000036285810000*** | −0.0000057441580000*** | 0.0000027855210000*** | 0.0000021501070000*** |
| | (0.0019379510000000) | (0.0000076746970000) | (0.0000000013184340) | (0.0000000014871350) | (0.0000001098473000) | (0.0000001015717000) |
| listed_count | −0.0148222500000000 | | 0.0000015926010000*** | | 0.0000030228860000*** | |
| | (0.0092338950000000) | | (0.0000000029407350) | | (0.0000005234023000) | |
| followers_count | | −0.0000268025300000*** | | 0.0000000045930810*** | | 0.0000000053589140*** |
| | | (0.0000000703115000) | | (0.0000000000053970) | | (0.0000000009305554) |
| hashtag_count | 1,480.2340000000000000*** | 1,486.5950000000000000*** | −0.1147696000000000*** | −0.1335951000000000*** | −0.0266094600000000 | −0.0245791600000000 |
| | (486.2257000000000000) | (1.9290520000000000) | (0.0001807942000000) | (0.0001805569000000) | (0.0275610300000000) | (0.0255311800000000) |
| hashtag_count2 | −132.9042000000000000* | −133.4696000000000000*** | 0.0071918160000000*** | 0.0097631630000000*** | 0.0020462090000000 | 0.0019218600000000 |
| | (77.8153700000000000) | (0.3087518000000000) | (0.0000307183900000) | (0.0000293161200000) | (0.0044108600000000) | (0.0040863410000000) |
| hashtag_sentiment | −102.3613000000000000 | −101.5614000000000000*** | 0.0406143000000000*** | 0.0448946800000000*** | 0.0172637300000000 | 0.0175941900000000 |
| | (356.5331000000000000) | (1.4146420000000000) | (0.0001281750000000) | (0.0001290223000000) | (0.0202096300000000) | (0.0187228300000000) |
| hashtag_sentiment2 | −21.3514500000000000 | −20.6422800000000000*** | 0.0058029650000000*** | 0.0066689690000000*** | −0.0016768320000000 | −0.0018304160000000 |
| | (83.3739700000000000) | (0.3307977000000000) | (0.0000276592600000) | (0.0000274938100000) | (0.0047259590000000) | (0.0043781270000000) |
| favourites_count | −0.0064924130000000* | −0.0064009580000000*** | 0.0000000038431430*** | 0.0000000300409000*** | −0.0000006634270000*** | −0.0000006308826000*** |
| | (0.0033426190000000) | (0.0000132478500000) | (0.0000000009590976) | (0.0000000009574004) | (0.0000001894709000) | (0.0000001753368000) |
| statuses_count | 0.0112412600000000*** | 0.0110702500000000*** | −0.0000007516136000*** | −0.0000007113500000*** | −0.0000000539127800 | 0.0000000141569000 |
| | (0.0029927080000000) | (0.0000118640800000) | (0.0000000010473000) | (0.0000000010571970) | (0.0000001696369000) | (0.0000001570205000) |
| s1_res | | 0.9999926000000000*** | | 0.0150136600000000*** | | 0.0000096708080000*** |
| | | (0.0000292825100000) | | (0.0000031008300000) | | (0.0000010184860000) |
| Constant | 4,186.3790000000000000* | 4,208.6040000000000000*** | 10.0713900000000000*** | 10.1454200000000000*** | 9.4648150000000000*** | 9.4674540000000000*** |
| | (2,405.9300000000000000) | (9.5451420000000000) | (0.0005906766000000) | (0.0005866992000000) | (0.1363749000000000) | (0.1263281000000000) |
| Observations | 18,420 | 18,420 | 18,420 | 18,420 | 18,420 | 18,420 |
| $R^2$ | 0.6464678000000000 | 0.9999944000000000 | | | | |
| Adjusted $R^2$ | 0.6453318000000000 | 0.9999944000000000 | | | | |
| Log Likelihood | | | −128,607,830.0000000000000000 | −121,441,487.0000000000000000 | −188,294.3000000000000000 | −187,622.0000000000000000 |
| $\theta$ | | | | | 2.4391840000000000*** (0.0213319000000000) | 2.6420170000000000*** (0.0243931500000000) |
| Akaike Inf. Crit. | | | 257,215,780.0000000000000000 | 242,883,096.0000000000000000 | 376,708.5000000000000000 | 375,366.1000000000000000 |
| Residual Std. Error | 24,188.2900000000000000 (df = 18360) | 95.9731300000000000 (df = 18359) | | | | |
| F Statistic | 569.0346000000000000*** (df = 59; 18360) | 54,979,525.0000000000000000*** (df = 60; 18359) | | | | |
| *Note:* | | | | | | *p<0.1; **p<0.05; ***p<0.01 |

Table 4: The OLS, Poisson, and Negative Binomial Models

Our analysis indicates that Tweets from unverified accounts have a higher likelihood of being retweeted as compared to Twitter verified accounts. Tweets that contain photos also are more likely to be retweeted. Across all models, it was found that the more the tweet was "liked", the more likely it was for it to be retweeted. This could be due to a "spill over" effect, where friends of users may observe that they have "liked" a tweet, and hence are now likely to react in a more elevated manner than their friend. The number of friends also was in a positive linear association with retweet count. This would indicate that the greater level of effort that a user spends in building connections, the greater the chance their tweets will be retweeted. The number of followers also had a positive effect on retweets.

The number of users that a user follows, their chance of being retweeted is greater than the contribution to retweets from the number of followers. That is, it is more beneficial for an account to connect with other users by "following" than it is to seek "followers". In addition, the number of statuses was in a positive linear association with retweet count. The number of hashtags was found to be linearly negative with retweets, however, it was also found that in the curvilinear model, it is positive, although this was not statistically significant. The opposite observation was found in the hashtag sentiment. The more positive/negative in sentiment of the hashtag, the less likely it would be retweeted. We also notice that the number of favourites the user liked negatively impacted the number of retweets. This is interesting, since it runs counter with our other finding of the friend's variable. These observations are unique and interesting and warrant additional research into associated causalities.

**Discussion**

The exploratory regression model seems to indicate a few findings regarding posting strategy so as to increase the probability of retweets. First, the model indicates how the user can strategically place themselves in a position to increase their chances of retweets of their own tweets by manipulating certain elements of their own account. First, the model indicates that the user should be relatively new. The longer the user is on twitter, the less likely they are to have their tweets retweeted. Therefore, after a few years, users may want to consider refreshing their page by deleting their old account and creating a new one. Next, we find that the users should be proactive by expanding their networks, but not overly "interacting" in those networks. They should place more of their effort in "following" others, rather than trying to build their own

"following". If future research further supports this result, then this certainly would be an interesting contribution to the current stream of literature, since this would contradict traditional thinking.

What also is interesting is the negative association between a user's activity in "liking" other tweets and the level of retweets for their own tweets. This would indicate that a user should be proactive in "following" other pages, but not to overly interact with those pages by "liking" their tweets. Rather, the user should spend more of their time tweeting on their own page. The number of total tweets that a user has was found to be positively associated with the probability of obtaining higher retweet counts, and hence another strategy is for the user to "tweet as much" as possible. Furthermore, non-verified accounts were found to be significantly different in retweet counts from verified. This is very interesting and needs additional research to conclude if users are more likely to retweet from non-verified accounts than verified ones. This has significant implications for Twitter and for users because if this is found to be true, then more users may prefer to not verify their account with Twitter.

The sentiment scores were mixed in our analysis, and so it would be difficult to generalize any findings along this line. Textual analytics lexicons are never perfect and there may inherent challenges based on the lexicons we had used. The negative association between tweet text length and retweet count would indicate that users should focus more of their efforts at producing short tweets. While the hashtag count had low significance, the coefficients indicate that fewer (very low) or more (very high) hashtags would result in higher retweets. That is, if hashtags are to be leveraged, then the user should either user as few as possible, or as many as possible. If they use a standard number (1-2 hashtags), their chances seem to not be as likely. In addition, hashtag sentiment was shown to be negatively associated with retweets, although not significantly. This would indicate that if hashtags are used, they should be neutral in sentiment. Furthermore, users should consider including a photo in their tweets, as well as tweeting from a mobile device rather than from a desktop or another device. These were found to be positively associated with retweets.

## Limitations

While our research had revealed some interesting patterns within the Twitter data, it was not without fault. First and foremost, our data collection efforts were mainly centered around a single month of collection. Unfortunately, the Twitter API only allows a limited timeframe of access to information, and so a more thorough data collection effort over a longer period of time would be necessary. In addition, the cross-sectional approach to our study is also a limitation of this research. The scope of this research was within only a small amount of time in a first attempt to ascertain some patterns that may be typical within Twitter data. However, these patterns may be time-dependent. Furthermore, the construct of "viralness" is inherently dependent on time. While our future research will address this limitation, it was none-the-less outside the scope of this research. In addition, our empirical model was centered around exploratory analysis rather than a theoretical justification of a conceptual model. Our purpose was primarily to determine factors that can be used to describe and explain the "viral" construct from a single moment in time.

## Conclusion

Our research has the potential for providing unique insights for understanding the viral nature of tweets which can be used by organizations to develop policies to support the development of best practices in disseminating information using Twitter. Future research can certainly be guided by our unique findings. While we had limitations within our data collection and analysis methodology, we were able to extract very useful information regarding the association of various user and tweet attributes and the nature of what we believe to be "viralness". As a result of our research, we propose additional research questions that would be addressed within future research – How would time-dependent effects be used to further explain and measure the construct of viral tweets? Does there exist a better lexicon, which is currently not being leveraged which would more appropriately describe the overall emotional sentiments of tweets? How would time-dependent effects on sentiment affect viralness. Furthermore, do there exist any interaction effects among the variables we have identified, and if so, what is the theoretical justification underlying these? In addition, we have not studied our construct of viralness from a network perspective. Which network structural characteristics impact viralness? Would these interact with the variables we have identified in this research?

In summary, we have described the construct of "viralness" from a static lens of cross-sectional analytics. Using a unique approach, we have confirmed prior research that two main types of variables affect this construct - namely, the user characteristics and the specific tweet characteristics. We have found some counter-intuitive results, such as non-verified accounts and shorter-aged accounts have a positive association with tweet popularity. These associations may be worthwhile to study in future research so as to better understand the complex behavioral dimensions that would explain these observations. Last, we have identified the overall posting strategy that would increase the likelihood of tweets being retweeted in the form of limited text and strategic use if hashtags. The managerial implications of this is that social media marketing strategy can be partially constructed from innovative decision support systems programmed to leverage the associations.